# Table-top setup for investigating the scintillation properties of liquid argon

T. Heindl[1], T. Dandl[1], A. Fedenev[2], M. Hofmann[3], R. Krücken[1], L. Oberauer[3], W. Potzel[3], J. Wieser[4], and A. Ulrich[1]

[1]Physik Department E12, Technische Universität München, James-Franck-Str. 1, 85748 Garching, Germany
[2]GSI Helmholtzzentrum für Schwerionenforschung GmbH, Planckstr. 1, 64291 Darmstadt, Germany
[3]Physik Department E15, Technische Universität München, James-Franck-Str. 1, 85748 Garching, Germany
[4]excitech GmbH, Branterei 33, 26419 Schortens, Germany

**Abstract.** The spectral and temporal light emission properties of liquid argon have been studied in the context of its use in large liquid rare-gas detectors for detecting Dark Matter particles in astronomy. A table-top setup has been developed. Continuous and pulsed low energy electron beam excitation is used to stimulate light emission. A spectral range from 110 to 1000 nm in wavelength is covered by the detection system with a time resolution on the order of 1 ns.

## 1. Introduction

Liquid rare-gases have attracted great attention as an efficient scintillator material for large scale particle and partly also gamma ray detectors [1-4]. The scintillation light of liquid argon including its time structure is for example used in detectors which search for Weakly Interacting Massive Particles (WIMPs) [5-7] or is proposed as a background veto in the GERDA detector [8]. Very little is known, however, about the detailed scintillation properties of this material in particular with respect to its emission spectrum. To the best of our knowledge, the last detailed studies of the spectroscopy of liquid rare-gases were performed more than 20 years ago and published in references [9-12], which are nowadays frequently cited. In these references, however, only a very limited spectral range is shown, hindering the assignment of different temporal structures. On this background we decided to adapt a technique which has been developed for vacuum ultraviolet light sources and gas kinetic studies to study the emission of liquid rare-gases focusing on the needs of recent detector developments.

An electron beam of 12 keV particle energy is used as the excitation source to induce scintillation light in liquid rare-gas samples. The key technology is to use extremely thin ceramic membranes through which the electrons are sent into a cryogenic target cell. The energy loss in these entrance foils is only 15% of the particle energy [13]. The basic concept of this technique has been described in detail in the context of its usage with gas targets and VUV light sources in previous publications [14-17]. Here it is shown that it can also be successfully applied to cryogenic targets filled with liquid rare-gases. Measurements for liquid argon have been performed. Measurements for the heavier rare-gases krypton and xenon are in preparation.

Based on our background in light source development, we were also interested if the electron beam induced scintillation of liquid argon can be used to build an efficient source in the VUV wavelength range. Brilliant VUV sources working in those wavelength regions can for example be used for single photon ionization in the field of mass spectrometry [18, 19].

## 2. Experiment

### 2.1 Conceptual design

An overview over the central part of the experimental setup is shown as a schematic drawing in figure 1. The rare gases are condensed into a target cell which is mounted in the center of a 63 mm diameter vacuum cube (CF63). A 12 keV electron beam is produced in a regular CRT (cathode ray tube) type electron gun in the vacuum outside the target cell and coupled into the cell through the thin membrane mentioned above. Fluorescent light is observed using $MgF_2$ windows and analyzed with a VUV monochromator. A liquid nitrogen Dewar from a germanium detector was modified to provide the cooling of the target cell. The gas system consisted of a closed cycle with a rare-gas purifier, a gas reservoir, and a metal bellows pump for circulating the gas through the system. The warm part of the gas system was connected with the target cell via a heat exchanger.

Experimental details concerning the target cell, the membranes, the operation of the electron gun, gas handling, the optical design, and the data taking are described in the following paragraphs. Experimental results for purified liquid argon have been published in a recent letter paper [20]. Results presented in this publication focus on the influence of impurities on liquid argon emission spectra.

### 2.2 Design of the target cell

The target cell was made of copper. A sectional view and a photograph of the cell are shown in figure 2 and figure 3,

respectively. Figure 2 has been adapted from a technical drawing and shows the conceptual design of the target cell drawn to scale. The cell was a cylinder with an inner diameter of 17 mm and an outer diameter of 50 mm. The entrance foil for the electron beam was mounted on a flange extending 15.7 mm into the cell reducing the distance to the $MgF_2$ window to 2.3 mm. This distance was chosen to allow comparative measurements using argon in the gas phase at room temperature down to a pressure of ~500 mbar. The range of 12 keV electrons in argon at atmospheric pressure is on the order of 1.2 mm. For a detailed discussion of the range and the shape of electron beam excited volumes in gases see reference [21]. The density of liquid argon is 1.4 g/cm³, and therefore 800 times larger than the density of gaseous argon under normal conditions. This leads to a glowing layer of about 1.5 µm thickness in liquid argon. A simulation of the spatial distribution of power deposited in liquid argon verifying this estimate is shown in figure 4.

thermal contact. This copper rod was clamped onto a second copper rod which was mounted to the inner volume of a thermally insulated liquid nitrogen container ("Dewar"). Gas in- and outlet were provided by two copper pipes with 3 mm inner diameter extending from the cell in vertical direction. They were both wound into spiral shape and soldered together forming a heat exchanger. The pipes were soldered into the cell and connected to stainless steel tubes in a vacuum feedthrough flange to connect them with the outside part of the gas system. The $MgF_2$ window was pressed against the cell's body using an indium ring as a gasket. The silicon nitride membrane was glued onto an invar (FeNi36 alloy) flange using epoxy glue [UHU Endfest 300]. This invar flange was then pressed against the cell's body, again using an indium gasket.

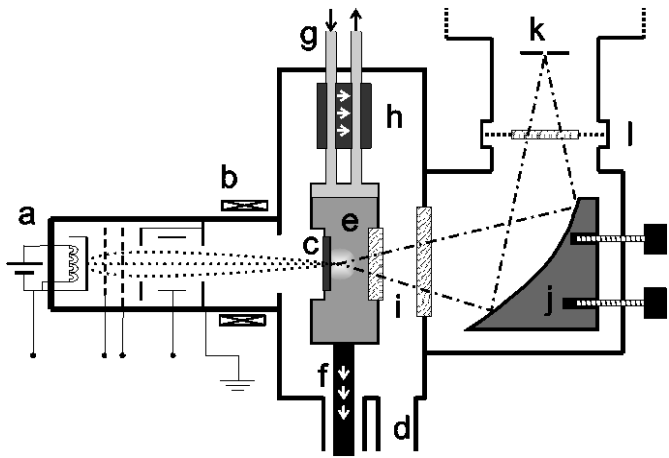

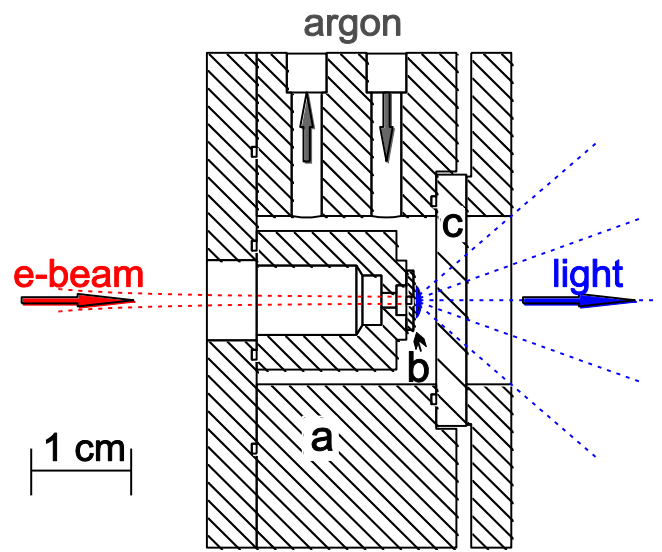

**Fig. 1.** Schematic drawing of the experimental setup: (a) electron gun with heated cathode, extraction and focusing gids; (b) magnets for beam steering; (c) 300 nm thick $SiO_2/Si_3N_4$ entrance membrane. The cell and electron gun volume (d) are evacuated with a turbo molecular pump. The liquid rare gas cell (e) (see also figure 2) was made of copper and was cooled via a connecting rod (f) attached to a liquid nitrogen Dewar. The liquid rare gas was continuously circulated (condensed and evaporated) in the cell via the gas in- and outlet pipes (g) which are in thermal contact with each other forming a heat exchanger (h). The fluorescence was observed through a pair of $MgF_2$ windows (i) separating the liquid rare gas from the vacuum surrounding the cell and the vacuum in the optical system, respectively. Light was collected by an adjustable elliptical mirror (j) and refocused onto the entrance slit of the VUV monochromator (k). A filter wheel (l) was used to suppress higher diffraction orders while recording spectra, at wavelengths larger than 200 nm.

**Fig. 2.** A sectional view adapted from a technical drawing of the liquid-argon cell is shown. The cylindrical cell (a) was manufactured from copper. The electron beam entered the inner volume of the cell through a tubular workpiece manufactured from invar. The $SiO_2/Si_3N_4$ membrane (b) was glued to the right end of this invar cylinder to separate the vacuum from the liquid-argon region. The electron beam can pass this membrane without significant energy loss. The fluorescence light was observed using a $MgF_2$ window (c).

The cell was held in position in the middle of a CF63 cube by a copper rod. This rod was screwed into the cell pressing an indium layer between the end of the rod and a flattened section of the cell's wall for providing good

"Super-insulation" in the form of 1.5 µm thick Al-coated mylar foil was loosely wrapped around all cryogenic parts since thermal radiation would contribute significantly to the thermal load with approximately 10 W. The average electron beam power, in comparison, deposited in the cell is only on the order of 10 mW. The access for the electron beam and the light output were of course not covered by super-insulation.

## 2.3 The membrane

The key technology which allowed us to build the table top setup described in this paper is a very thin ceramic membrane for coupling low energy electron beams from the vacuum into gases and liquids [14, 22]. These membranes [Fraunhofer- Institut für Zuverlässigkeit, München] are 300 nm thick and consist of silicon nitride or a combination of silicon nitride and silicon oxide for compensation of internal stress. They are manufactured by solid state technology (chemical vapor deposition) on the surfaces of both sides of a silicon wafer. Then one side is structured and the wafer material is etched away in the areas where free standing membranes should be produced. The ceramic material acts as an etch-stop in this process. The membranes used here have a size of 0.7 × 0.7 mm$^2$ on a 5 × 5 mm$^2$ wafer-section. For the cryogenic target with a temperature down to -190 °C used in this study it was important to test whether the membranes would withstand the temperature change from room temperature to this low temperature region, still tolerating the stress of sending a significant electron beam current through them. They passed a test of immersing them into liquid nitrogen easily. However in order to match the expansion coefficients it was necessary to glue them on invar metal instead of copper for obtaining proper vacuum sealing. Beam currents up to, and beyond the point where liquid argon is starting to evaporate could be applied as will be shown below.

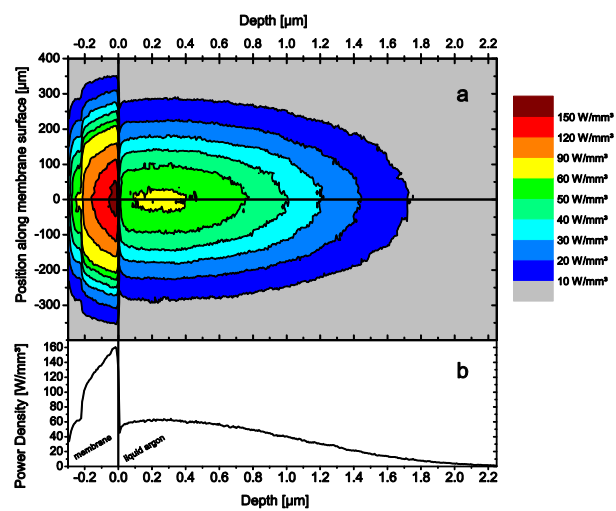

**Fig. 4.** *A Monte Carlo simulation of the deposited power density by 12keV primary electrons using an 1.5 μA electron beam for excitation is shown. The simulation was performed using the program CASINO V2.42 [24]. Plot (a) shows data for the power density in a plane containing the beam axis. Plot (b) shows the power density deposited in a small volume around the beam axis. The maximal power density deposited in the liquid is about 60 Wmm$^{-3}$ while the power density deposited in the membrane (negative values in depth) reaches values of 160 Wmm$^{-3}$. One can see the two layer composition of the membranes, due to the different densities of silicon oxide (left side) and silicon nitride [13].*

Energy loss of the electrons in the membrane and the energy distribution of electrons exiting the membrane are described in reference [13]. The energy loss is on the order of 15%. Furthermore, for a detailed description of the energy deposited in the scintillating material, backscattering of electrons in the membrane and the target material has to be considered. Details are shown in the next paragraph and in figure 4.

### 2.4 Electron gun operation

Applying a stabilized, computer controlled power supply [Optimare EPU], an electron beam with 12 keV particle energy was used for the excitation of the liquid argon. The electron beam was used either in a continuous mode or in a pulsed regime with pulses as short as 100 ns. It is possible to generate electron beam pulses in the 1-5 ns range with the electron guns used. The technique for short pulse generation is available in our laboratory [23] and will be adapted to the liquid rare gas setup in the future.

Evaporation of argon due to the energy deposited by the electron beam can "contaminate" the liquid with gaseous argon. Therefore, the electron beam current had to be limited to 1.5 μA (continuous) in experiments where it is the aim to study the emission from undisturbed liquid argon. Higher beam currents led to evaporation of the gas which can be observed by a strong appearance of the atomic 4p-4s ArI lines in the red and near infrared as can be seen in figure 5 (see paragraph 2.6 for measurement technique). Observing the emission region by eye, one can clearly see the evaporation in liquid-argon at higher beam currents, due to bubble production at the membrane surface. The absence of the ArI lines was also used in the experiments with pulsed excitation as an indication for undisturbed conditions. All measurements presented in the following, were performed below the threshold current for evaporation.

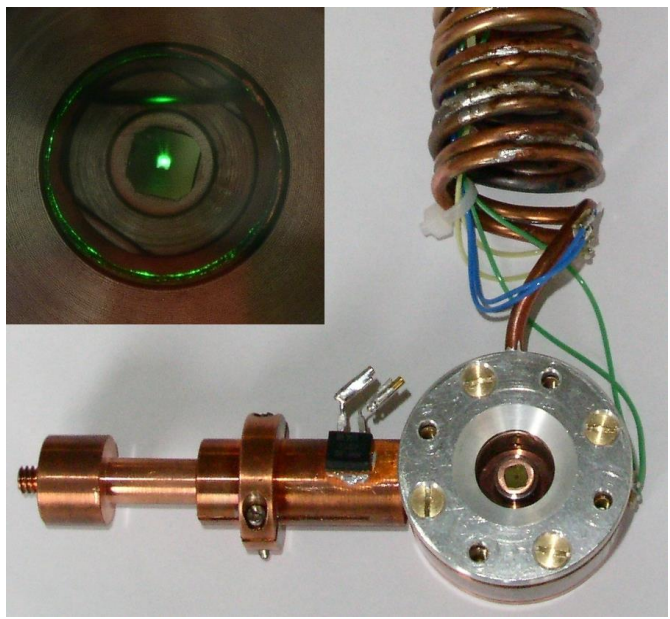

**Fig. 3.** *Photograph of the liquid argon cell prior to its installation in the setup. It shows the attached copper rod (left-hand side) which established the thermal connection to the Dewar. The heat exchanger (on top) was built from two copper pipes soldered to each other. To fit into a CF63 pipe they were wound in a corkscrew way with approximately 5 cm diameter. The insert shows a photograph of the emission from liquid argon. The greenish colour is due to an oxygen impurity [32, 33] in the liquid argon since the purification was not turned on while taking this picture. The greenish glow above the central area is caused by the reflection of scintillation light at the argon gas-liquid boundary. One can also see some reflected light from the indium gasket, used to seal the MgF$_2$ window.*

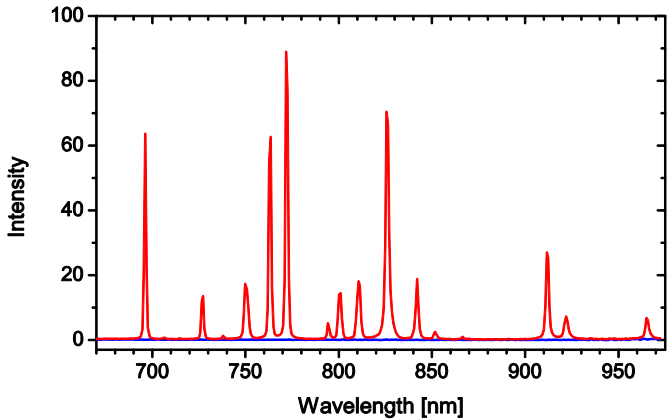

*Fig. 5.* A spectrum of electron beam excited liquid argon in the red and near infrared spectral region is shown. No features can be observed in this spectrum when the electron beam current was limited to a maximum of 1.5 µA (cw) (blue line). Using higher beam currents led to partial evaporation of the liquid argon in the vicinity of the membrane. This excited, gaseous argon emits its typical atomic 4p-4s ArI lines, which can be observed in the spectrum (red line, recorded at 3 µA excitation current). This test was used to find the threshold current for evaporation of liquid argon in the setup.

A Monte Carlo simulation was performed using the program CASINO V2.42 [24]. Figure 4 shows simulation results for the deposited power in the membrane and the liquid argon using an 1.5 µA electron beam for excitation. The accuracy of the results given by CASINO simulations have been studied in earlier publications with respect to spatial power deposition in gases [21] and the transmission through the silicon nitride membranes [13]. It is interesting and important to note for the application of modeling scintillation detectors in which only individual projectiles interact with matter that the interaction of the electrons with the target is still via individual collisions despite the very small volume of the beam excited liquid. This can be estimated using the following arguments: The Monte Carlo simulation (see figure 4) gives a value of about 60 W per mm³ of maximal power density deposited in the liquid. A value of $W_i$=23.6 eV for the average energy needed to cause one ionization in liquid argon is reported in reference [2]. This leads to a production rate of $1.6 \times 10^{19}$ secondary electrons per second and mm³. The slowest time constants which can be observed in the fluorescence after pulsed excitation are on the order of 20 µs (fluorescence of xenon impurity, see figure 4 of reference [20]). After that time interval essentially all excitation within the volume has decayed. Multiplying the production rate with 20 µs results in a density of secondary electrons of $3.2 \times 10^{14}$ mm$^{-3}$. This value is $6.6 \times 10^4$ times smaller than the density of atoms in condensed argon ($2.1 \times 10^{19}$ mm$^{-3}$).

An analogous calculation results in a very similar number for the maximum density of secondary electrons in liquid argon, when the pulsed excitation was used.

## 2.5 Gas handling and cooling

Gas purity is a key issue for obtaining efficient vacuum ultraviolet excimer emission from the rare gases both in the gas phase [25, 26] and the liquid [11, 27, 28]. The liquid argon cell was connected via the heat exchanger to a gas system at room temperature with a rare gas purifier [SAES Getters, MonoTorr® Phase II, PS4-MT3]. A buffer volume of ~3 l which was baked before the measurements was installed in the gas system and used as a reservoir of purified argon. A metal-bellows pump was used for circulating the gas continuously through the system. Prior to the measurements the evacuated and baked gas system was filled with 1300 mbar argon (purity: 4.8 [Linde] or 6.0 [Westfalen AG]). The pressure was measured using a precision membrane capacity gauge [MKS Baratron 390H 1000]. The argon was purified in the system and the light emission of gaseous argon could be observed in the target cell. After about 1 h of purification no impurity lines, besides the xenon line (see below) could be seen in the emission spectrum. Cooling and condensation of the argon in the target cell was initiated by filling liquid nitrogen coolant into the Dewar. When reaching a temperature of approximately 86 K, which was used for most of the measurements, the target cell (volume: ~2 cm³) was filled completely with liquid argon. The temperature of the target cell could be monitored with three Pt100 platinum resistance thermometers glued onto the target cell at different positions on its outer surface. The temperature of the cell could be adjusted by the gas flow through the system and a high-power heating resistor soldered to the copper rod connecting the cell and the cold part of the Dewar. The Pt100 temperature sensors were calibrated during the condensation of argon, matching the readout of the high precision pressure gauge to literature values of the boiling point of argon at various pressures [29]. Details of this calibration are shown in figure 6.

## 2.6 Optics design

Light produced in the target cell was decoupled with a VUV-transparent MgF$_2$ window and focused onto the entrance slit of a f=30 cm VUV monochromator [McPherson 218] (see figures 1 and 2 for details). An additional MgF$_2$ window was installed between the vacuum around the target cell and the vacuum inside the monochromator so that the monochromator could be vented without breaking the vacuum around the target cell. An elliptical mirror with Al-MgF$_2$ coating was used as the optical element between the cell and the monochromator. Using a filter wheel, it was possible to position a sapphire or a glass filter in the light pass to suppress higher diffraction orders of the monochromator grating (1200 lines mm$^{-1}$, blaze 150 nm, wavelength resolution: 0.6 nm at 100 µm slit-width) while recording the wavelength spectra. Venting the monochromator and removing the elliptical mirror enabled us to observe the fluorescing argon with a wide range (UV-NIR) spectrometer [Ocean Optics QE65000], or to take photographic pictures of the fluorescing region, respectively, as shown in the insert of figure 3. An important improvement of the VUV light detection system with respect of the early measurements [9, 11] was the usage of a photomultiplier (PMT) with a MgF$_2$ window and a VUV sensitive multi-alkali cathode mounted at the VUV monochromator instead of a wavelength shifter and a regular PMT.

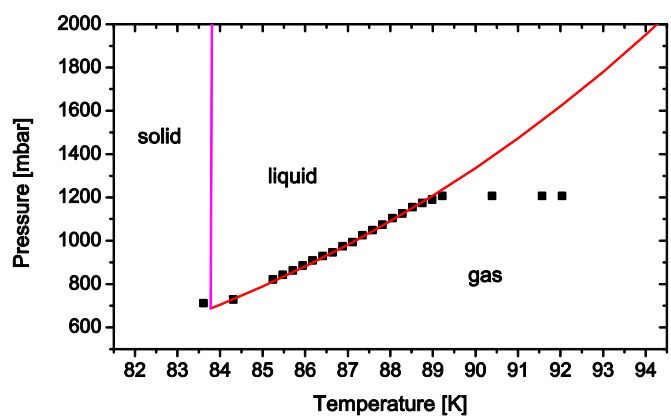

*Fig. 6.* *Phase diagram of argon. Full lines show literature data [29]. The squares show measured values of argon pressure versus temperature in the setup during cooling and condensation. Using this plot an initial offset in the temperature measurement (Pt100 sensor) of -1.4 K was found. This constant was then used to correct the data obtained from the Pt100 temperature sensors.*

To obtain a response function of the optical system in the VUV-region (115-250 nm) spectra of electron beam excited gaseous argon at room temperature were recorded with the target-cell. These spectra were compared with data obtained at identical conditions in a setup which was built to study the emission of gaseous argon [25], and which had been calibrated using a deuterium lamp [Cathodeon D$_2$-Lamp, traceable to: type V03, serial No: V0282, calibrated by PTB calibration mark PTB-065402] with known emission spectrum. Besides a small error due to the described transfer of the calibration between two setups the potential variation in the shape of spectra can be assessed by listing the wavelength dependence of the expanded uncertainty (coverage factor: 2) in the absolute flux of the calibration lamp in different wavelength intervals (116.0-120.4 nm: 14%, 120.6-122.6 nm: 36%, 122.8-170.0 nm: 14%, 172.0-250.0 nm: 7%). The wide range spectrometer (QE65000) was calibrated using a halogen lamp with known emission spectrum [LOT 100 W, type LSB20, serial No. LSK173] (expanded uncertainty, coverage factor: 2, 200-295 nm: ≤ 12%, 295-400 nm: 9.4%, 400-800 nm: 8.2%, 800-1100 nm: 8.6%). Both calibrations are based on the spectral radiance. The intensity axes in the wavelength spectra shown below are therefore proportional to the light output power per wavelength interval (W/nm).

## 2.7 Data recording system

The PMT signals obtained in photon counting mode were processed using a constant fraction discriminator (CFD). The pulses were registered in a counter module for continuous excitation. The photomultiplier had a dark count rate of 100 Hz while signal count rates in the most intense emission features were on the order of 20 kHz which is well below PMT pileup conditions. In the pulsed excitation mode the time delay between a trigger pulse which starts the electron beam and a photomultiplier pulse from a detected photon was measured with a time-to-amplitude converter [Ortec 566]. The output signal of the module has an amplitude proportional to the time difference between the start and stop pulses. The amplitude was measured with a 13 bit analog-to-digital converter [Ortec AD413A]. The time measurement system was calibrated using a precision pulse generator [Stanford Research Systems DG535]. Data from the analog-to-digital converter as well as from the counter were transferred to a personal computer using a CAMAC-to-PC interface controller [Wiener CC32].

## 3. Results and Discussion

### 3.1 Spectral emission from liquid argon

Emission spectra of purified liquid argon were recorded. The result is shown in figure 7a. Details about the emission from pure liquid argon have been presented in ref. [20]. The spectra are dominated by a broad emission feature at about 126.8 nm, which is the analogue of the 2nd excimer continuum in the gas phase. Spectra in the gas phase could also be recorded with the setup described here. The 126.8 nm emission intensity in the liquid was compared with measurements performed in the gas phase at 1300 mbar. The intensity in the liquid was found to be 59% of the intensity in the gas phase. Since the absolute emission intensity in the gas phase had been measured in an earlier experiment [25], it is possible to calculate the florescence efficiency for liquid argon in our setup. From this comparison we derive a value of $(1.9\pm0.3)\times10^4$ VUV-photons emitted per MeV deposited. This result is by a factor of 2 lower, than a literature value for (relativistic) electrons [2]. A possible explanation for this discrepancy may be a space charge induced electric field produced by the electrons sent into the target. Decreasing fluorescence efficiency with increasing electric field in liquid rare gases due to reduced recombination is a well known effect [2]. Dedicated measurements of the fluorescence efficiency and its field dependence will be performed in the future.

## 3.2 Spectral emission from impurities in the liquid argon

Similar to the gas phase, the VUV emission spectrum of liquid argon is very sensitive to impurities. Heavier rare-gases such as xenon or krypton and impurities like oxygen, nitrogen, and carbon play an important role [20]. The latter three impurities can be removed in our system by the rare-gas purifier before and during the measurements. Removing a rare-gas impurity, in contrast, is difficult and cannot be done with the purifier used here. Since our gas system had been used with xenon before, spectra of the first argon fillings showed intense emission structures originating from xenon. Most pronounced was the emission of the xenon resonance line at 148.9 nm [11]. Note that for this publication the wavelength of this emission has been slightly corrected (in reference [20] we reported 149.1 nm). After month-long repetitive flushing and operation with pure argon from the bottle the amount of xenon could be reduced, drastically. Spectra with different amounts of xenon impurity recorded during this preparation phase of the experiment are shown in figure 7b and 7c. Note, that for none of these spectra, xenon was added deliberately. This is why we cannot give a quantitative estimate about the amount of xenon in the system, while these spectra were recorded. The spectrum in figure 7a was recorded after the rare-gas purifier had been replaced by a new one of the same type. It still shows light emission caused by traces of xenon.

The spectra (figure 7) show a clear increase of the xenon resonance line at 148.9 nm with increasing xenon content. Note that the xenon resonance line has shifted from 147.0 nm in the gas phase to 148.9 nm in liquid argon [11] while the linewidth of the emission increases from <0.6 nm (resolution of spectrometer) to about 1.5 nm (FWHM). That means the energy of the xenon resonance line is decreased by about 0.11 eV when the xenon atom is in a liquid argon matrix. A modification of the shape of 2nd continuum of argon and the appearance of the 2nd continuum of xenon can also be observed at the highest xenon concentrations measured (figure 7c). This effect is well known from the gas phase [30]. Note that the analogue of the 3rd continuum [31] disappears with the appearance of the xenon features. For higher xenon concentration the 2nd continuum emission of the xenon excimer around 175 nm becomes the dominant emission in the spectrum, as reported in references [10, 11].

To test the influence of other impurities, we recorded spectra, without using the gas purification system. Argon directly from the gas-bottle (purity: 4.8), was filled and condensed in our nominally clean gas system. The changes in the VUV spectrum can be seen in figure 5 of reference [20]. A new broad emission structure around 197 nm which can be tentatively attributed to an oxygen impurity [32, 33] appears in the spectrum, while the intensity of the analogue of the 2nd continuum (around 127 nm) decreases. The emission at the longer wavelengths is also altered as can be seen in figure 8. In the visible spectral range all broadband emission features lose intensity, while the peak at 557 nm, which is caused by the emission of an oxygen impurity [32] gains intensity. Therefore, the emission of unpurified liquid argon appears as a greenish glow to the human eye (insert in figure 3), while purified liquid argon shows a whitish glow.

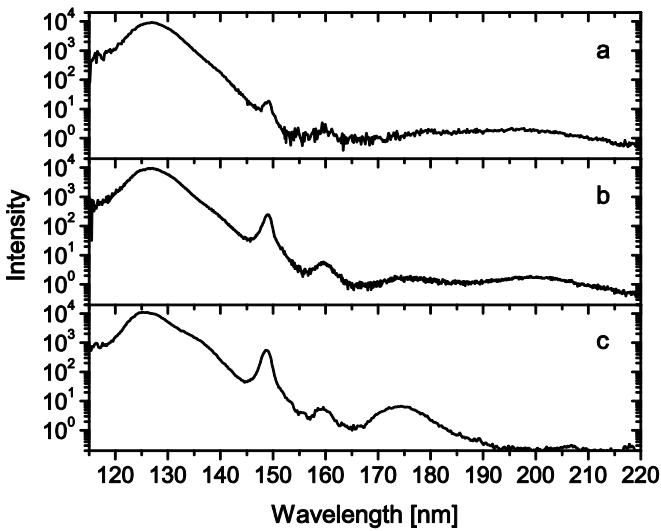

*Fig. 7.* Liquid argon spectra, at different levels of xenon impurity. With increasing xenon impurity (from a to c) in liquid-argon, the emission structure at 148.9 nm increases. At a high level of xenon admixture (c) the argon continuum is modified in such a way that it extends to longer wavelengths. Furthermore, another broad feature appears in the spectra around 175 nm which is the analogue of the 2nd continuum emission of xenon excimer light in the gas phase while the 3rd continuum emission of argon (around 200 nm) [31] disappears.

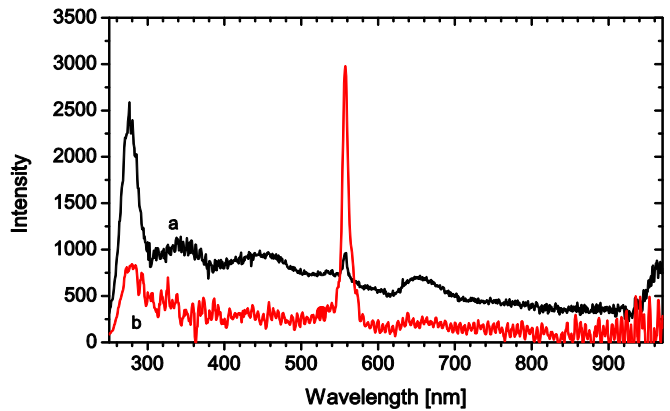

*Fig. 8.* Emission spectra of purified (black line, a) and unpurified (red line, b) liquid argon. The peak at 557 nm is caused by the emission from an oxygen impurity [32]. When the purification system is not used in our setup, the intensity of this peak increases drastically, while all other emission features in the spectral range shown here lose intensity

## 3.1 Time resolved measurement of the emission from impurities

Time evolution of the light emission at specific wavelengths can be recorded with the monochromator and the photomultiplier in the setup presented here. An important issue regarding scintillation detectors is the question, how the time-structure of scintillation light changes due to impurities in liquid-argon. In most liquid-argon scintillation detectors the wavelength-shifted VUV emission is detected and used for particle discrimination and identification [5-8, 34]. This VUV emission is normally attributed to the 2$^{nd}$ continuum, only. The decay of the 2$^{nd}$ continuum emission follows two time constants which are interpreted in terms of the different lifetimes of the singlet and triplet state of the excimer molecules. We have measured those lifetimes in purified liquid-argon. For the long lifetime (triplet) we measured a value of (1300±60) ns. Due to limitations in pulsing the electron beam, we can so far only give an upper limit of <6.2 ns for the short lifetime (singlet). Details on these measurements and a comparison with literature values can be found in reference [20].

For the low levels of xenon impurities in our setup (according to all spectra in figure 7) we have found no influence on the time-spectrum (time dependence of light intensity) of the 2$^{nd}$ continuum emission of liquid-argon. But, as mentioned above, xenon leads to additional emission features in the spectra. In detectors, where wavelength integrating detection techniques are used, this leads to a measured time-spectrum in which several time structures of different emission processes are superimposed. The interpretation of those time-spectra is difficult for that reason. Therefore it is important to study the time structure of emission features caused by impurities separately in defined wavelength intervals. In the case of a xenon impurity, we recorded a time-spectrum at 148.9 nm, the central wavelength of the xenon resonance line in liquid-argon. The time-spectrum of the xenon impurity can be found in figure 4 of reference [20]. The intensity of its emission grows to a maximum ~2 μs after the excitation pulse and decays to zero about 20 μs after the excitation pulse.

We also recorded time-spectra in liquid-argon, while the gas-purification was turned off. It was found that the temporal emission of the analogue of the 2$^{nd}$ argon continuum changes drastically. Due to quenching by impurities, the lifetime of the triplet state is significantly reduced. With an impurity level according to the spectrum shown in figure 5 of reference [20] it reduces from 1300 ns to 342 ns (figure 9). The time-spectrum of the emission around 197 nm which was attributed to an oxygen impurity in the paragraph above is also shown in figure 9.

## 4. Outlook

We presented a setup for studying the scintillation properties of liquid argon in a broad wavelength interval (110 – 1000 nm). Excitation was performed using low energetic electron beams. Wavelength spectra and wavelength-resolved time-spectra have been recorded of all major emission features of purified liquid argon and of some emission features caused by prominent impurities.

Experiments using different ion beams for the excitation of liquid argon in a similar setup are currently performed. Time- and wavelength-spectra obtained with excitation by different projectiles probing the particle discrimination capabilities of liquid argon scintillators will be compared in a forthcoming publication.

A recent publication [8] has shown that adding xenon deliberately to a liquid-argon detector in which a wavelength shifter is used, can improve its particle discrimination capabilities. Therefore, a detailed study of the wavelength and time resolved scintillation of liquid argon with well defined admixtures of xenon in our setup may help to find an optimized mixture for scintillation detectors with respect to particle discrimination.

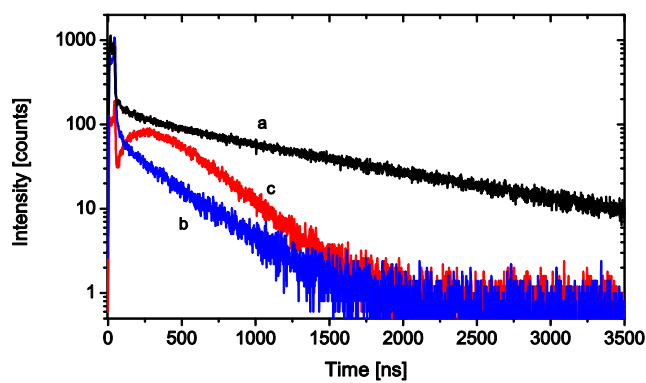

**Fig. 9.** *Time-spectrum emitted from the 2$^{nd}$ continuum (126.8 nm) of liquid argon following pulsed electron beam excitation. The lifetime of the triplet state was measured to be ~1300 ns in purified conditions (black line, a, see also reference [20]). Due to quenching by impurities the lifetime is reduced to ~342 ns when the liquid argon was not purified (blue line, b). Furthermore unpurified liquid argon showed an additional emission feature around 197 nm in our wavelength spectra (see figure 5 in reference [20]), which may be caused by an oxygen impurity [32, 33]. A time-spectrum emitted from the emission feature at 197 nm is also shown here (red line, c).*

The purity of the condensed rare-gas is a key issue in all liquid rare-gas scintillation detectors [2, 6, 27, 28, 35]. Purity in these detectors is normally monitored, by measurements of free electron lifetimes in the condensed rare-gas, or by measuring the decay time of the slow component (triplet state). But as shown in this publication, the latter technique is not very useful to detect small amounts of xenon for example. Our method of electron beam excitation combined with an observation of wavelength resolved scintillation light might be an alternative method for online monitoring and identification of impurities in large volume liquid rare-gas detectors.

Minor modifications to our setup, will enable us to study the scintillation of condensed heavier rare-gases in the future. Because of its widespread use in operational or proposed experiments for WIMP detection [35-37], the investigation of scintillation properties of condensed xenon will be a subject for future investigations.

## Acknowledgments

We thank Teresa Marrodán-Undagoitia for her useful comments after reading a draft of this paper. This work has been supported by the Maier Leibnitz Laboratory Munich (MLL), the German Ministry of Education and Research (BMBF), contract No. 13N9528, and the Deutsche Forschungsgemeinschaft DFG (Transregio 27: Neutrinos and Beyond).